\documentclass[twocolumn]{aastex631}

\usepackage{amsmath}
\usepackage{subfigure}
\usepackage{amsmath}
\definecolor{blue-violet}{rgb}{0.30, 0.1, 0.89}

\begin{document}

\title{Polarized Emission of Intrabinary Shocks in Spider Pulsars from Global 3D Kinetic Simulations}

\author[0000-0002-9545-7286]{Andrew G. Sullivan}
\affiliation{Kavli Institute for Particle Astrophysics and Cosmology, Department of Physics, Stanford University, Stanford, CA 94305, USA}

\author[0000-0001-5186-6195]{Jorge Cort\'es}
\affiliation{Astronomy \& Astrophysics Department, University of California Santa Cruz, Santa Cruz, CA, 95064, USA}
\affiliation{Department of Astronomy and Columbia Astrophysics Laboratory, Columbia University, New York, NY 10027, USA}

\author[0000-0002-1227-2754]{Lorenzo Sironi}
\affiliation{Department of Astronomy and Columbia Astrophysics Laboratory, Columbia University, New York, NY 10027, USA}
\affiliation{Center for Computational Astrophysics, Flatiron Institute, 162 5th avenue, New York, NY, 10010, USA}



\begin{abstract}
In spider pulsar systems, a relativistic intrabinary shock forms when the pulsar wind collides with the massive outflow driven off the pulsar's low-mass stellar companion. The shock is a site of non-thermal particle acceleration, likely via shock-driven magnetic reconnection, and produces synchrotron emission. These shocks are among the few systems in which global scales can be reasonably captured with kinetic simulations, enabling first-principles particle acceleration and emission studies. We perform the first global 3D kinetic simulations of spider pulsar intrabinary shocks and predict their polarized emission properties. We report emission spectra, light curves, and polarization patterns as a function of the stripe-averaged magnetic field,  cooling strength, and viewing inclination. At $90^\circ$ inclination and for a low stripe-averaged magnetic field, we reproduce the double peaked light curve observed in spider systems. We predict a significant polarization degree $\gtrsim15\%$, which monotonically increases with the stripe-averaged field strength. Our results can be applied to and tested by forthcoming X-ray polarization observations of spider pulsars.
\end{abstract}



\section{Introduction} \label{sec:intro}

In tight binary systems with orbital periods $P_b\lesssim1$ day, millisecond pulsars (MSPs)  known as spider pulsars irradiate and ablate their low-mass companion stars. This drives off a massive wind---{with mass flux $\sim10^{-10}$ M$_\odot$ yr$^{-1}$}---from the companion surface \citep{1988Natur.334..225K,1988Natur.334..684V, 2013IAUS..291..127R, 2019Galax...7...93H, 2024ApJ...974..315S}, which collides with the pulsar wind to form an intrabinary shock (IBS). Spider pulsars are subdivided into black widows with companion mass $M_c< 0.1$ M$_\odot$ and redbacks with $M_c\approx 0.1-0.4$ M$_\odot$.  In redbacks, which typically have higher stellar wind momentum fluxes, the IBS wraps around the pulsar, while in black widows {with lower stellar wind momentum fluxes}, the IBS wraps around the companion \citep{2016ApJ...828....7R, 2017ApJ...839...80W, 2019ApJ...879...73K}. 

The IBS represents a potential site of non-thermal electron and positron ($e^\pm$) acceleration. 
The high-energy particles emit synchrotron X-rays \citep{2019ApJ...879...73K, 2021ApJ...917L..13K}. The light curves are orbitally modulated, with two peaks associated with beamed emission from relativistic particles traveling along the shock surface \citep{2019ApJ...879...73K, 2021ApJ...917L..13K, 2024ApJ...974..315S}. 
The X-ray spectra in spiders are observed to be extraordinarily hard, with photon indices $\Gamma\sim1-1.5$ \citep{2012ApJ...756...33C, 2014ApJ...793L..20R, 2014ApJ...795...72L, 2019ApJ...879...73K}  and in some cases $\Gamma<1$ \citep{2025ApJ...984..146S}. 
Shock-driven magnetic reconnection of the striped pulsar wind likely produces the non-thermal particle spectra needed to explain the hard X-ray spectra \citep{2012ApJ...756...33C, 2014ApJ...793L..20R, 2014ApJ...795...72L, 2019ApJ...879...73K, 2011ApJ...741...39S, 2021ApJ...908..147L,  2022ApJ...933..140C, 2024MNRAS.534.2551C, 2025MNRAS.tmp..266C, 2024ApJ...974..315S, 2025ApJ...984..146S}.

Past fluid numerical simulations of intrabinary shocks, most of which are aimed at modeling high companion mass $\gamma$-ray binaries,  provide a description of the shock geometry and effects of orbital motion \citep{2008MNRAS.387...63B, 2012MNRAS.419.3426B, 2012A&A...546A..60L, 2015A&A...577A..89B, 2015A&A...581A..27D, 2019MNRAS.490.3601B, 2024A&A...690A..75G}, but lack the physics required for understanding the emission signatures. More recently, global kinetic Particle-in-Cell (PIC) simulations of binary pulsars have become possible \citep{ 2024A&A...689A.251R, 2024ApJ...973..147Z}, including IBS simulations in 2D \citep{2022ApJ...933..140C, 2024MNRAS.534.2551C, 2025MNRAS.tmp..266C}. The latter capture the basic black widow geometry and provide first-principles predictions for the IBS non-thermal emission. Due to the 2D geometry, the companion is (unrealistically) modeled as a cylinder. This affects the shock geometry and reconnection properties. The IBS structure is inherently 3D given that the companion wind is roughly spherical. A 3D IBS model enables observable predictions at inclinations outside the orbital plane. Furthermore, predicting the polarized emission --- one of the next frontiers of spider pulsar observation --- requires accurate 3D magnetic field modeling \citep{2023IBSPolarization}. 

In this paper, we perform 3D PIC simulations of an IBS which wraps around the companion star. We report first-principles predictions for particle spectra and polarized synchrotron emission.  We describe the simulation setup in Sec. \ref{sec:simsetup}, particularly its extension to 3D. In Sec. \ref{sec:spectra}, we calculate the non-thermal particle energy and synchrotron emission spectra derived from the simulations. In Sec. \ref{sec:LCs}, we present the predicted synchrotron light curves and phase modulated polarization information. We discuss the astrophysical implications of our findings and conclude in Sec. \ref{sec:conclusion}.

\section{Simulation Details}
\label{sec:simsetup}
We extend the 2D setup of \citet{2022ApJ...933..140C, 2024MNRAS.534.2551C, 2025MNRAS.tmp..266C} to 3D in the PIC code \texttt{TRISTAN-MP} \citep{2005AIPC..801..345S}. We use the Vay pusher to integrate the particle equations of motion \citep{2008PhPl...15e6701V}.  
The plasma skin depth $c/\omega_p$, {where $\omega_p$ is the electron plasma frequency,} is resolved with 2 grid cells while the numerical speed of light is $c=0.45$ cells/timestep. As in \citet{2022ApJ...933..140C, 2024MNRAS.534.2551C, 2025MNRAS.tmp..266C}, we inject a plane parallel $e^\pm$ pulsar wind moving along $-\hat{x}$ from the rightmost boundary of the box, which starts next to the companion and propagates towards $+\hat{x}$ \citep{2011ApJ...741...39S, 2022ApJ...933..140C}. 
The leftmost $x=0$ boundary absorbs both particles and fields. The $y$ and $z$ boundaries are periodic, and the box in both $y$ and $z$ is 4080 cells wide. We initialize the plane-parallel pulsar wind magnetic field {using a Harris-like profile} \citep{1962NCim...23..115H, 2025ARA&A..63..}
\begin{equation}
\label{eq:Harrissheet}
    B_y(x, t)=B_0 \tanh\left\{\frac{1}{\Delta}\left[\alpha+\cos{\left(\frac{2\pi(x+\beta_0ct)}{\lambda}\right)}\right]\right\},
\end{equation}
where $B_0$ is the field amplitude, $\lambda=2\pi P_sc$ is the striped wind wavelength with $P_s$ as the pulsar spin period, $\Delta$ is proportional to the width of the current sheet separating the stripes, $\beta_0=(1-1/\gamma_0^2)^{-1/2}$ is the initial velocity of the wind (specified by the bulk Lorentz factor $\gamma_0$), and $\alpha=2\left<B_y\right>_\lambda/(B_0+\left<B_y\right>_\lambda)$ quantifies the field averaged over one striped wind wavelength, $\left<B_y\right>_\lambda$ \citep{2011ApJ...741...39S, 2022ApJ...933..140C}. We specify the wind magnetization $\sigma=B_0^2/4\pi n_0\gamma_0m_ec^2$, with $n_0$ as the cold wind $e^\pm$ number density. The cold wind temperature is initially set {to be non-relativistic} with $kT_c/m_ec^2=10^{-4}$ {(or equivalently $T_c\approx6\times10^5$ K)}. In the hot current sheets, the $e^\pm$ temperature is set by balancing the thermal pressure in the sheet with the magnetic pressure in the stripe, i.e. $kT_h/m_e c^2=\sigma/2\eta$, where $\eta$ is the sheet to stripe particle density ratio.

In our simulations, we set $\sigma=10$,  $\gamma_0=3$, $\lambda=100\,c/\omega_p$, $n_0=1$ particle per cell, $\Delta\times\lambda=10\pi\,c/\omega_p$ ({so that the current sheet width is $5\,c/\omega_p$}), and $\eta=3$. This sets the characteristic Lorentz factor for reconnection accelerated particles $\gamma_\sigma\equiv\gamma_0\sigma=30$ whose Larmor radius is $r_{L, \sigma}=\sqrt{\sigma}\,c/\omega_p\approx3\,c/\omega_p$. The values of $\gamma_0$ and $\sigma$ we choose are smaller than those expected in realistic spider systems. \citet{2024MNRAS.534.2551C} show that different $\gamma_0$ merely leads to an overall shift of all energy scales, while magnetization does not substantially affect the results as long as $\sigma\gg1$.

We require a higher value of $\Delta$ than \cite{2022ApJ...933..140C} to prevent the onset of reconnection ahead of the termination shock in 3D. The arbitrary choice of $\eta$ affects the energy of the particles initialized in the current sheets. We are most interested in acceleration and emission of particles from the cold pulsar wind, which we will focus on in the rest of the paper. We find that the obtained downstream spectra of the cold wind particles are nearly insensitive to the choice of $\eta$. 

The companion is modeled as a sphere with radius $R_c$ which sources an ultra-relativistic isotropic wind with bulk Lorentz factor $\gamma_w=60$ and number density $n_w=1\, n_0$ at the companion surface. {A dense non-relativistic wind, as the real companion wind is expected to be, is extremely difficult to model in a relativistic PIC simulation, since most of the computational power would be spent on evolving companion wind particles.} We aim to study only the physics of the pulsar wind, so this ``unrealistic" companion wind model is meant only to halt and shock the pulsar wind \citep{2022ApJ...933..140C, 2024MNRAS.534.2551C}. See \citet{2022ApJ...933..140C} for a more detailed discussion.

As in \cite{2025MNRAS.tmp..266C}, we include synchrotron cooling using the reduced Landau-Lifshitz radiation reaction force \citep{2016CoPhC.204..141V}. The cooling strength is set by the synchrotron burnoff limit $\gamma_{\mathrm{rad}}$ defined by
\begin{equation}
    eE_{\mathrm{acc}}=\frac{4}{3}\sigma_T\gamma_{\mathrm{rad}}^2 \frac{B_0^2}{8\pi},
\end{equation}
where $E_{\rm acc}$ is the accelerating electric field, and $\sigma_T$ is the Thomson cross section.
In reconnection, $E_{\mathrm{acc}}\approx\beta_{\mathrm{rec}} B_0$ where $\beta_{\mathrm{rec}}\approx0.1$ is the relativistic reconnection rate \citep{2025ARA&A..63..}. This provides a parameterization for the ratio of the radiation reaction force to the Lorentz force.

We vary two parameters across our simulations: the stripe-averaged magnetic field set by $\alpha$ {in eq. \ref{eq:Harrissheet}}, and the cooling strength set by $\gamma_{\mathrm{rad}}$. {Varying $\alpha$ has the effect of changing the width of one magnetic stripe relative to the two neighboring ones, without changing $\sigma$ or $B_0$.} We use $\alpha=\left\{0.0,\,0.1,\,0.3,\, 0.5\right\}$. The value of $\alpha$ ranges from 0 to 1 with latitude above the pulsar spin equator over an angle equal to the pulsar obliquity \citep{1999A&A...349.1017B}. In MSP binaries, the orbital angular momentum is typically assumed to be aligned with the pulsar spin, so that $\alpha=0$ in the orbital plane and $\alpha\neq0$ at higher latitudes. For simplicity, we keep $\alpha$ the same at all latitudes in each run. We discuss the validity of this choice in Sec. \ref{sec:conclusion}. We perform simulations with $\gamma_{\mathrm{rad}}=\{30, 60, \infty\}$, spanning $\gamma_{\mathrm{rad}}/\gamma_\sigma=\{1, 2, \infty\}$. Our choices thus span the strong ($\gamma_{\mathrm{rad}}\sim\gamma_\sigma$) to weak ($\gamma_{\mathrm{rad}}\gg\gamma_\sigma$) cooling regimes. Observations place $\gamma_\sigma\gtrsim10^5$ and $\gamma_{\mathrm{rad}}\sim10^7$ \citep{2025ApJ...984..146S}.  Our fiducial companion has a radius of $R_c=100\, c/\omega_p$ and is centered at position $(x_c, y_c, z_c)=(750, 1020, 1020)\,c/\omega_p$. We evolve the simulations until they have reached a quasi-steady state which occurs at $t\gtrsim2800\,\omega_p^{-1}=28\,R_c/c$. We then compute the synchrotron emission properties for subsequent times as discussed in the following subsection.  We also test two smaller companion sizes: $R_c=50\, c/\omega_p$  centered at $(x_c, y_c, z_c)=(375, 510, 510)\,c/\omega_p$ and $R_c=70\, c/\omega_p$  centered at $(x_c, y_c, z_c)=(525, 714, 714)\,c/\omega_p$. We scale the box size in $y$ and $z$ proportionately, and evolve these to the same time in units of $R_c/c$ as the fiducial case.

\begin{figure*}
    \centering
    \includegraphics[width=\linewidth]{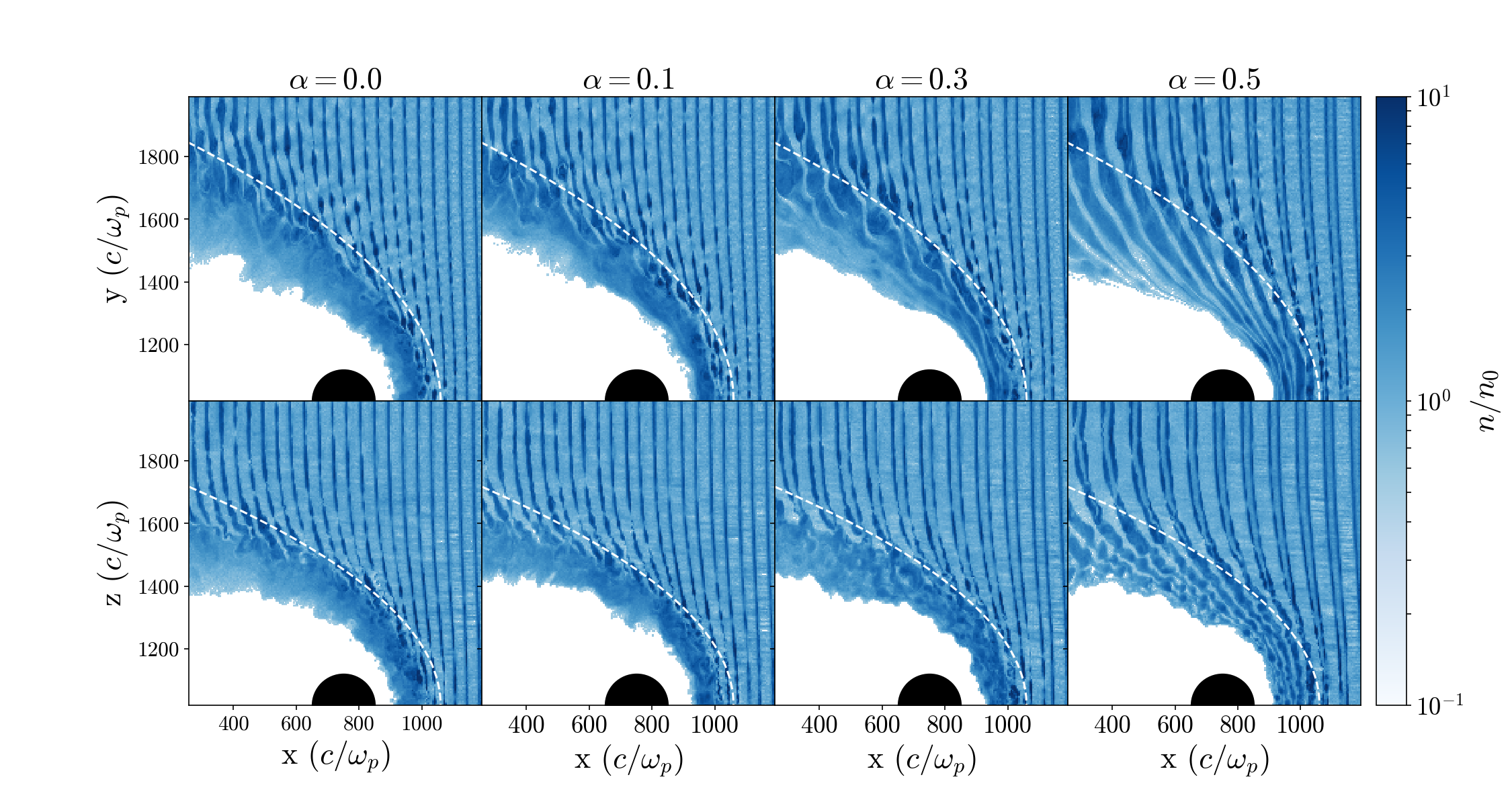}
    \caption{2D slices of the  pulsar wind particle number density in units of $n_0$ at $t=3375\,\omega_p^{-1}$ for the $R=100\,c/\omega_p$ cases without cooling, as a function of $\alpha$. The IBS downstream region is left of the dashed white curves. (Top) The $x$-$y$ plane at $z=1020\,c/\omega_p$. (Bottom) The $x$-$z$ plane at $y=1020\,c/\omega_p$ (i.e. the two slices are taken at the center of the companion). }
    \label{fig:density}
\end{figure*}

\subsection{Computing the Polarized Emission}
We compute both isotropic (i.e. angle-integrated) and viewing angle dependent polarized synchrotron emission. For the isotropic emission, we compute the synchrotron power \citep{1970RvMP...42..237B, 1979rpa..book.....R} from each particle that started in the cold wind and now resides in the IBS downstream. 
For orbitally modulated emission, we define a line of sight unit vector 
\begin{equation}
\boldsymbol{n}=\cos{\phi_B}\sin{i} \hat{x}+\sin{\phi_B}\sin{i} \hat{y} + \cos{i}\hat{z},
\end{equation}
where $i$ is the observer inclination angle defined with respect to the binary orbital angular momentum which is in the $\hat{z}$ direction, and $\phi_B$ is the azimuthal angle in the orbital plane. For each inclination, we divide the range of azimuthal angles into $N_{\mathrm{LOS}}=64$ normalized phases $\Phi_B$  (i.e. $\phi_B=2\pi\Phi_B-\pi/4$ so that $\Phi_B=0.25$ at superior conjunction). We compute the synchrotron intensity at a given $i$ and $\Phi_B$ by summing the contribution of the IBS downstream particles which originated in the cold wind whose velocities are within an angle $\theta<\pi/N_{\mathrm{LOS}}$ of $\boldsymbol{n}$. This method captures the effects of particle anisotropy and Doppler beaming.  

We choose our polarization angle $\psi$ to be defined so that $\psi=0$ along the direction of the binary orbital angular momentum projected in the plane of the sky $\hat{l}$ (i.e. $\hat{l}\perp\boldsymbol{n}$). The polarization angle is given by 
\begin{equation}        
\cos{\psi}=\hat{a}_{sky}\cdot\hat{l},
\end{equation}
where $\hat{a}_{sky}$ is a unit vector in the direction of the particle acceleration projected on the sky plane. Stokes parameters for the $j$th particle are thus computed as 
\begin{subequations}
    \begin{equation}
    Q_j=(P_{j,\perp}-P_{j,||})(\cos^2\psi-\sin^2\psi),
\end{equation}
\begin{equation}
    U_j=(P_{j,\perp}-P_{j,||})(2\cos\psi\sin\psi),
\end{equation}
\end{subequations}
where $P_{j,\perp}$ and $P_{j,||}$ are, respectively, the power radiated by the $j$th particle perpendicular and parallel to the local magnetic field \citep{1979rpa..book.....R}. The total $Q$ and $U$ at each $\Phi_B$ are computed in the same manner as the intensity. 

The Stokes $I$, $Q$ and $U$ computed in the simulation determine the polarization degree (PD)
\begin{equation}
    \Pi=\frac{\sqrt{Q^2+U^2}}{I},
\end{equation}
and the electric vector polarization angle (EVPA) 
\begin{equation}
    \psi=\frac{1}{2}\arctan_2\left(\frac{U}{Q}\right).
\end{equation}
With 3D simulations, we can properly compute emission for multiple viewing inclinations and lines of sight.

\section{Particle and Synchrotron Spectra}
\label{sec:spectra}
In Fig. \ref{fig:density} we show the pulsar wind particle number density at $t=3375\,\omega_p^{-1}$ (including  particles that started both in the cold wind and in the hot current sheets), for different values of $\alpha$. We find that the higher $\alpha$ cases do not form a strong shock, and the striped structure of the flow is somewhat preserved in the IBS downstream for $\alpha=0.5$. We define the IBS downstream as the region delimited by the dashed white lines (cross sections of a paraboloid). It is apparent that the shock is wider in the $x$-$y$ plane than in the $x$-$z$ plane, likely due to magnetic tension effects (the pre-shock field is along ${y}$ and the post-shock field is mostly in the $x$-$y$ plane). We describe the flow and field properties in more detail in a forthcoming paper. Below, we extract particle and synchrotron spectra from the IBS downstream region.

\begin{figure*}     
\includegraphics[width=\linewidth]{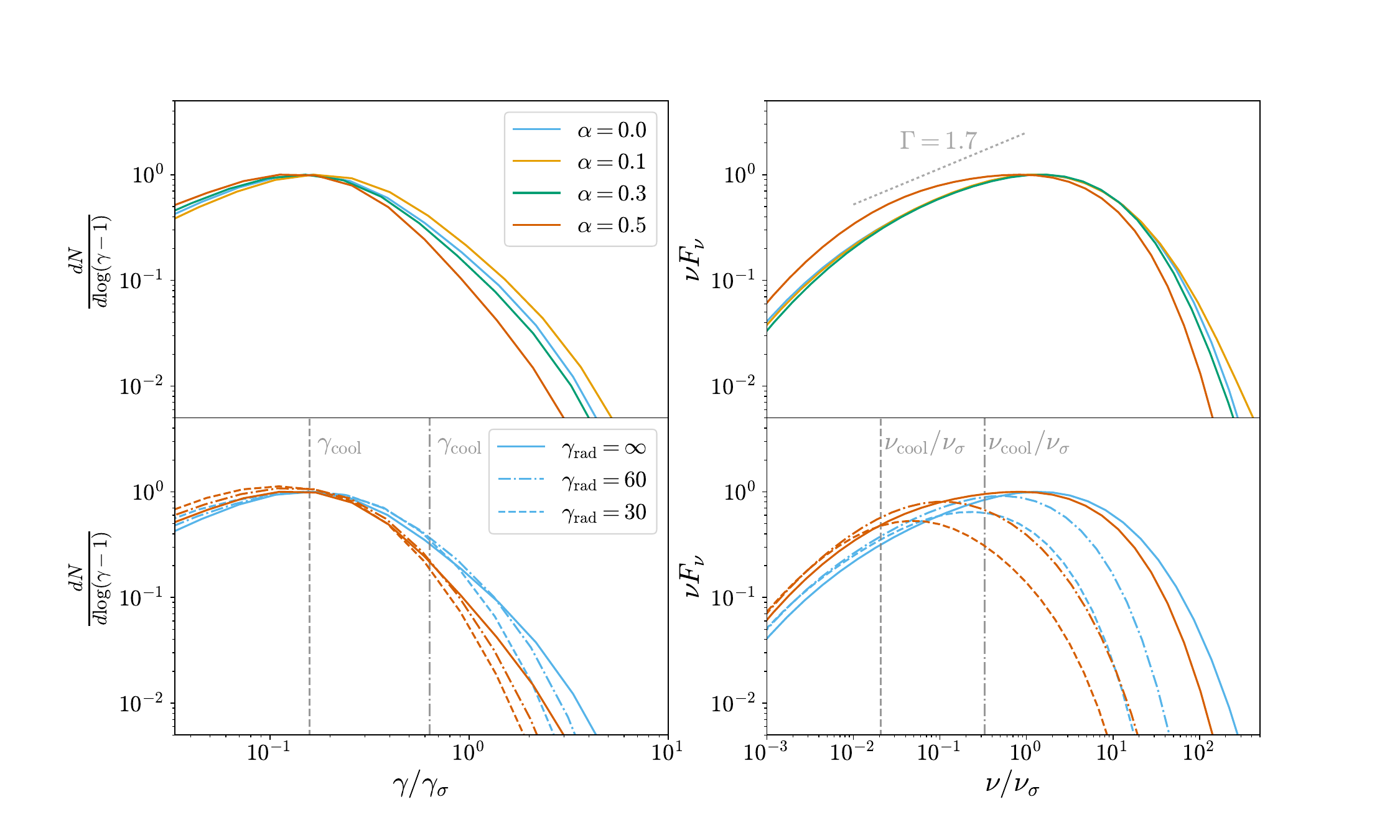}
    \caption{Particle energy spectra (Left) and isotropic synchrotron spectra (Right) for different $\alpha$ (Top) and different $\gamma_{\mathrm{rad}}$ (Bottom). Colors denote $\alpha$ while line styles denote $\gamma_{\mathrm{rad}}$, as illustrated in the legends. The top row refers to uncooled cases. In the bottom row, we show $\alpha=0.0$ (blue) and $\alpha=0.5$ (red) for different values of $\gamma_{\mathrm{rad}}$.}
    \label{fig:spectra}
\end{figure*}

In Fig. \ref{fig:spectra}, we show downstream particle energy spectra and isotropic synchrotron spectra for the range of $\alpha$ and $\gamma_{\mathrm{rad}}$  we explore.  As in 2D, very hard spectra are obtained for both particles and synchrotron photons. We begin discussing the cases without cooling (i.e. the top panels in Fig. \ref{fig:spectra}). Across all $\alpha$ cases, particles are accelerated up to $\sim \gamma_\sigma$ and beyond. The particle spectra are nearly the same for $\alpha\leq0.3$. The case $\alpha=0.5$ has a noticeably softer spectrum than the other cases. This is related in large part to differences in the shock structure (see Fig. \ref{fig:density}). 
When $\alpha=0.5$, a greater fraction of the upstream magnetic field is preserved across the shock, and the larger residual field prevents a strong shock from forming. Also, the fraction of upstream field energy available for particle energization is smaller for higher $\alpha$, leading to a steeper spectrum for $\alpha=0.5$.

The features of the particle spectra are imprinted on the synchrotron spectra. For all $\alpha$ cases, the synchrotron spectra peak near a characteristic frequency
\begin{equation}
    \nu_\sigma=\gamma_\sigma ^2\frac{eB_0}{2\pi m_ec},
\end{equation}
corresponding to emission from particles having a Lorentz factor $\gamma_\sigma$.  The synchrotron spectra are remarkably consistent for $\alpha\leq0.3$.  In the frequency range $10^{-2}<\nu/\nu_\sigma<1$, the spectrum can be modeled as a power law  $dN_\nu/d\nu\propto F_{\nu}/\nu\propto\nu^{-\Gamma}$ with $\Gamma\approx1.7$. The $\alpha=0.5$ case is noticeably softer than the other three cases in the same frequency range, which is a consequence of the softer particle spectrum.

We now discuss the cases with cooling (i.e. bottom panels in Fig. \ref{fig:spectra}). As compared to their uncooled counterparts, the cooled particle spectra soften above a characteristic Lorentz factor $\gamma_{\mathrm{cool}}$, which corresponds to the energy of particles that cool on a dynamical time (i.e. $ t_{\mathrm{cool}}\sim R_{\mathrm{curv}}/c$, where $ R_{\mathrm{curv}}\propto R_c$ is the shock curvature radius). The ratio of $\gamma_{\mathrm{cool}}$ to $\gamma_{\sigma}$ is
\begin{equation}
    \frac{\gamma_{\mathrm{cool}}}{\gamma_\sigma}=0.16\left(\frac{0.1}{\beta_{\mathrm{rec}}}\right)\left(\frac{\sigma}{10}\right)^{\frac{1}{2}}\left(\frac{200\,c/\omega_p}{R_{\mathrm{curv}}}\right)\left(\frac{\gamma_{\mathrm{rad}}}{\gamma_\sigma}\right)^2.
\end{equation}
Cooling signatures appear in the synchrotron spectra  above the cooling frequency $\nu_{\mathrm{cool}}=(\gamma_{\mathrm{cool}}/\gamma_{\sigma})^2 \nu_\sigma$. For realistic spider systems, $\nu_{\mathrm{cool}}$ is expected to fall in the X-ray band where the IBS emission is typically observed \citep{2025MNRAS.tmp..266C, 2025ApJ...984..146S}.  For $\alpha\leq 0.3$, the synchrotron spectra above the cooling frequency (i.e. from $\nu_{\mathrm{cool}}$ to $\nu_\sigma$) form a power law with index $\Gamma\approx 2$, which is softer than the $\Gamma\approx 1.7$ slope of corresponding uncooled cases. 
The bottom right panel of Fig. \ref{fig:spectra} shows that cooling effects are more pronounced for $\alpha=0.5$ than for cases of  $\alpha=0.0$ having the same $\gamma_{\rm rad}$. For $\alpha=0.5$, the synchrotron spectra cut off around $\nu_{\mathrm{cool}}$, rather than extending to $\nu_{\sigma}$.  This comes from a combination of two effects: first, as discussed before, the efficiency of magnetic field dissipation (and so, of particle energization) decreases with $\alpha$; second, the stronger post-shock residual field at higher $\alpha$ enhances synchrotron cooling losses.

\section{Polarized Light Curves}
\label{sec:LCs}
\begin{figure*}
    \centering
    \hspace*{-2cm}
    \includegraphics[width=1.2\linewidth]{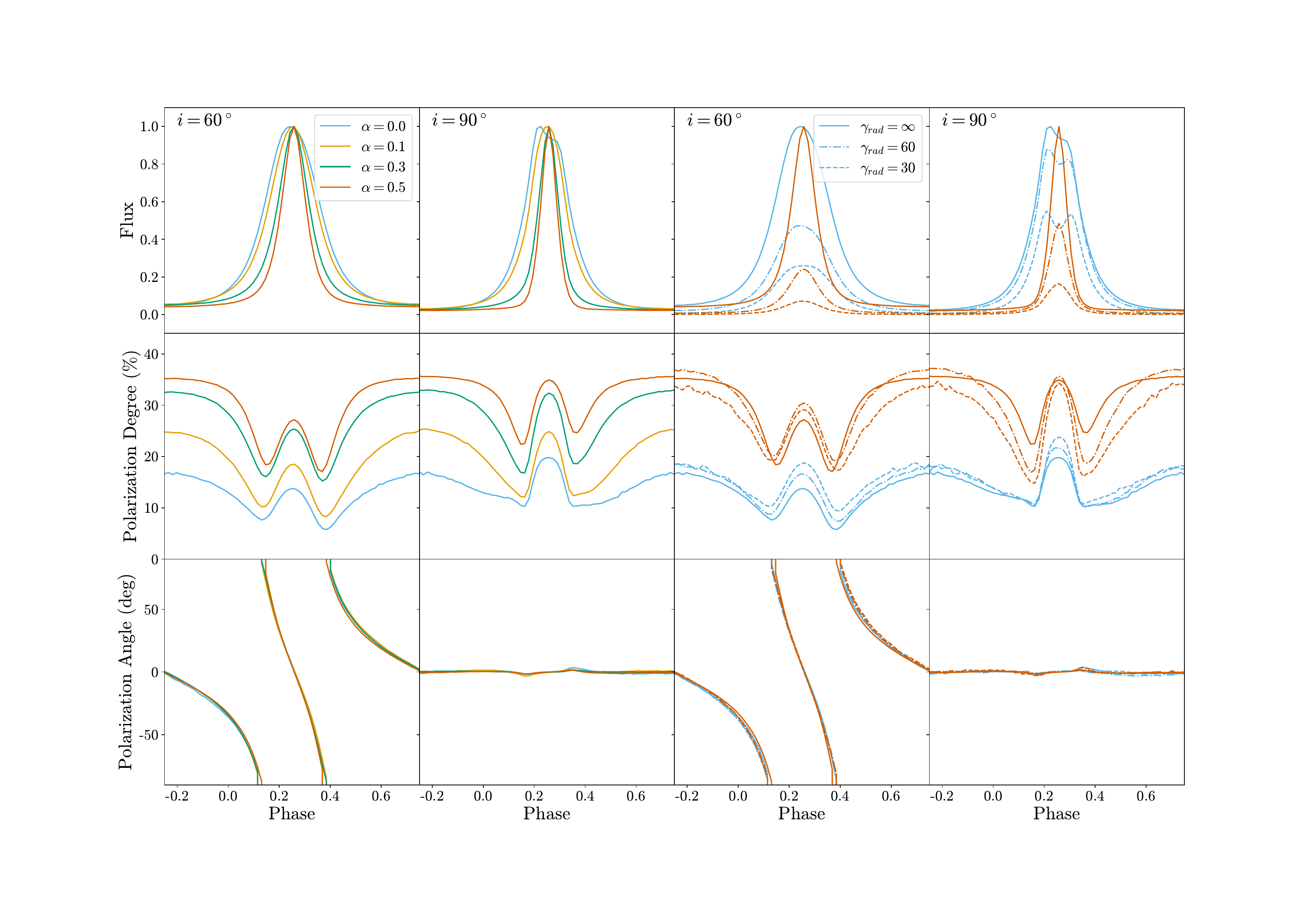}
    \caption{Orbital phase modulated flux, PD, and EVPA curves at $\nu=\nu_\sigma$. Phase $\Phi_B=0.25$ corresponds to the time when the observer is behind the shock along $-\hat{x}$ (i.e. pulsar superior conjunction). Color denotes different $\alpha$ while  line style denotes different $\gamma_{\mathrm{rad}}$ (same as in Fig. \ref{fig:spectra}). All light curves for a given $\alpha$ are normalized to their peak flux in the $\gamma_{\mathrm{rad}}=\infty$ case.}
    \label{fig:lcs}
\end{figure*}
In Fig. \ref{fig:lcs} we show light curves, as well as PD and EVPA patterns, for the $\alpha$ and $\gamma_{\mathrm{rad}}$ cases we explore. We show results for inclinations $i=60^\circ$ and $i=90^\circ$. We present synthetic observables measured at $\nu=\nu_\sigma$, but we remark that for $10^{-2}<\nu/\nu_\sigma<1$ (i.e. in the power-law portion of the spectrum) the light curve and polarization properties are nearly the same as the ones at $\nu=\nu_\sigma$. At $\nu>\nu_\sigma$,  the overall PD increases as expected for frequencies closer to the spectral cutoff, while the EVPA pattern does not change.

The emission is heavily inclination dependent, as expected \citep{2019ApJ...879...73K, 2023IBSPolarization}. We first consider the case $\alpha=0.0$. At $i=90^\circ$, the three $\gamma_{\mathrm{rad}}$ cases all display two peaks in the light curve, characteristic of emission from the two sides of the IBS, viewed just before and after superior conjunction (i.e. around $\Phi_B=0.25$). The two peaks become more prominent in the cooled cases because the  downstream region is thinner and thus the post shock flow is channeled over a narrower range of angles \citep{2025MNRAS.tmp..266C}, which enhances its Doppler beaming. At smaller $i=60^\circ$, the two peaks merge into one wide peak as emission from high latitudes (above the orbital plane) near the top of the shock contributes at $\Phi_B=0.25$. {The part of the shock above the companion thus contributes as much emission as the two sides (lying in the orbital plane) that produce the peaks when $i=90^\circ$.} A light curve with a single peak is observed in real spider systems believed to have small inclinations \citep{ 2020ApJ...904...91V, 2025ApJ...984..146S}, although typically for inclinations smaller than $i=60^\circ$ \citep{2019ApJ...879...73K}. We argue that the absence of two peaks at $i=60^\circ$ in our simulations comes from the fact that our companion radius is too small compared to the width of the post-shock flow. 

For higher $\alpha$ cases, the emission predominantly comes from the nose of the shock, where particle acceleration is most efficient and the compressed residual magnetic fields are the strongest. This causes their light curves to display a single peak even for $i=90^\circ$.

The PD and EVPA temporal patterns (but not the overall PD level) remain the same across the different $\alpha$ cases, but vary with inclination. At $i=90^\circ$, the polarization angle is constant over the orbit, corresponding to a polarization vector parallel to the binary angular momentum axis. The magnetic field, lying preferentially in the $x$-$y$ plane, is projected onto the sky plane perpendicular (making the polarization vector parallel) to the orbital angular momentum at all orbital phases. The PD drops just before and after $\Phi_B=0.25$, when the line of sight is nearly tangent to the post-shock flow direction. At these phases, multiple zones appreciably contribute to Doppler-beamed emission. The post-shock field orientation is not uniform across these zones, and the most strongly beamed zones have their magnetic field parallel to the line of sight \citep{2023IBSPolarization}, which reduces the PD.
At $i=60^\circ$, the magnetic field vector, which lies preferentially in the $x$-$y$ plane, rotates during the orbit when projected onto the sky plane, causing the EVPA to rotate as well. Particularly rapid polarization swings occur around superior conjunction for $i=60^\circ$ \citep{2023IBSPolarization}.

We conclude this section by highlighting two general trends. For a given $\alpha$ and $\gamma_{\rm rad}$, a lower PD is observed near $\Phi_B=0.25$ for $i=60^\circ$ than for $i=90^\circ$. As already mentioned, the post-shock field preferentially lies in the $x$-$y$ plane. For $i=90^\circ$, the polarization electric field of each emission zone is parallel to the orbital angular momentum; instead, for $i=60^\circ$, the polarization electric field direction varies across different zones, depending on how the post-shock magnetic field is oriented within the $x$-$y$ plane. This results in lower PD.

\label{sec:LCs}
\begin{figure}
    \centering
    \hspace*{-1.0cm}
    \includegraphics[width=1.2\linewidth]
    {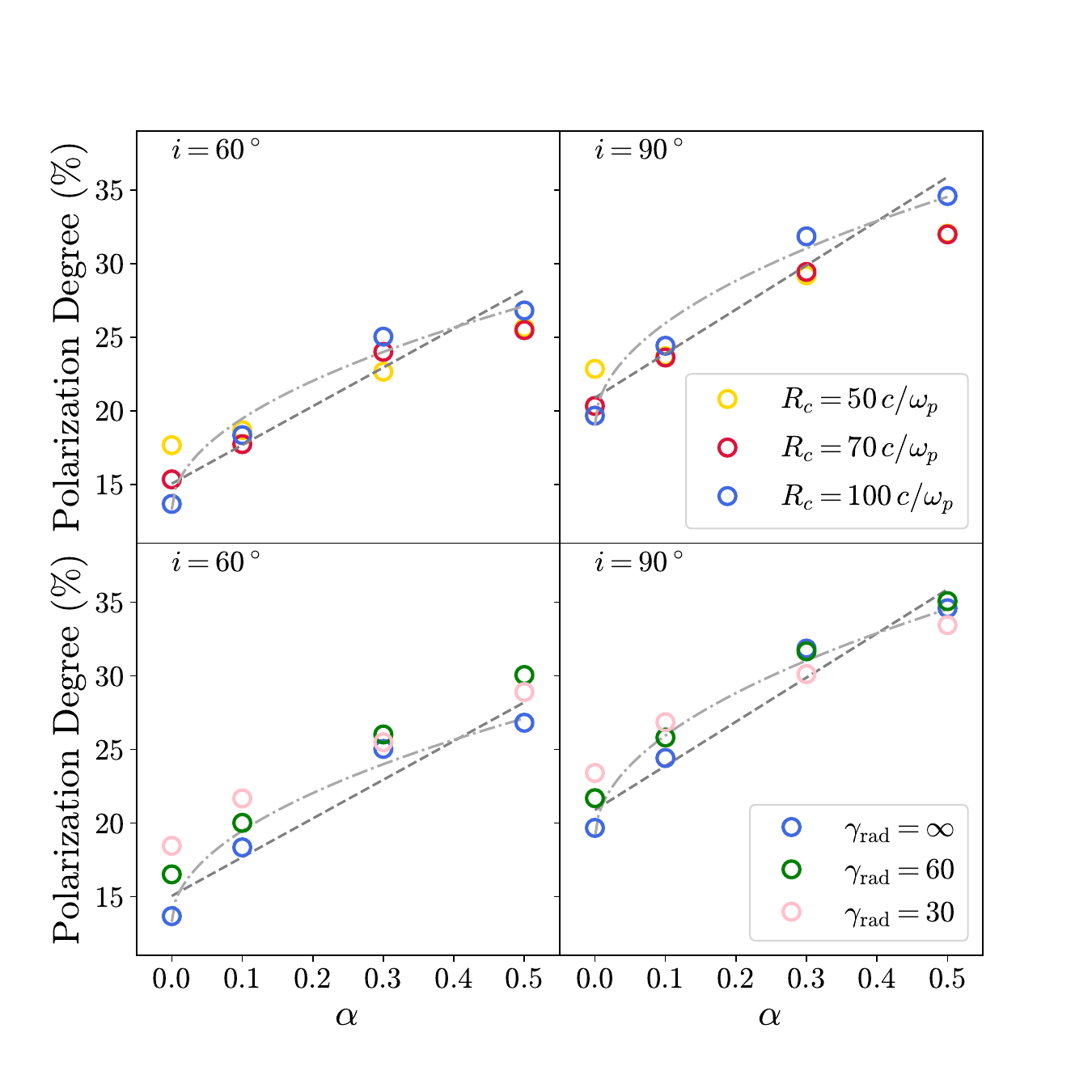}

    \caption{PD at/near flux maximum ($\Phi_B=0.25$) as a function of $\alpha$ for $i=60^\circ$ (left) and $i=90^\circ$ (right). Top panels show different companion sizes $R_c$ and bottom panels show different $\gamma_{\mathrm{rad}}$. Eq. \ref{eq:fit} with the parameters in table \ref{table:1} is plotted using dashed lines and dot-dashed lines for $b=1$ and $b=1/2$, respectively.}
    \label{fig:convergence}
\end{figure}
\begin{table}
\setlength{\tabcolsep}{10pt}
\begin{tabular}{cccc}
 
 \hline\hline
 $i$ & $b$ &  $A_i$ & $C_i$ \\
 \hline
$60^\circ$ & $1$ & 25\%& 15\% \\
$60^\circ$ & $1/2$ & $20\%$ & $13\%$ \\
$90^\circ$ & $1$ & 30\%& 21\% \\
$90^\circ$ & $1/2$ & $22\%$ & $19\%$ \\
 \hline\hline
\end{tabular}
\caption{Best-fit slope and constant offset values for eq. \ref{eq:fit}, which describes the dependence on $\alpha$ of the PD at flux maximum, for different $i$ and $b$. }
\label{table:1}
\end{table}
Another evident trend is the dependence of the overall PD level on $\alpha$, with PD increasing roughly proportionately to $\alpha$. In contrast, the temporal patterns of both PD and EVPA are practically independent of $\alpha$ as previously discussed. A stronger net field is preserved for higher $\alpha$ in the shock downstream, leading to higher PD. In contrast, lower $\alpha$ cases have stronger turbulent fields, leading to lower PD.

In Fig. \ref{fig:convergence}, we show the trend in PD with $\alpha$ at flux maximum with $i=60^\circ$ and $i=90^\circ$ for different companion sizes and $\gamma_{\mathrm{rad}}$. We fit the relationship between PD and $\alpha$ for a given $i$ as
\begin{equation}
\label{eq:fit}
    \text{PD}_{i}=A_i \alpha^{b}+C_i,
\end{equation}
with $b=1$ or $b=1/2$. We show the estimated values $A_i$ and $C_i$ for the fiducial uncooled $R_c=100\, c/\omega_p$ simulations in table \ref{table:1}.
These relationships vary little between different companion sizes and different $\gamma_{\mathrm{rad}}$. Stronger cooled cases generally show marginally higher PD, as expected due to their steeper spectra \citep{2023IBSPolarization}. The small and non-monotonic differences in PD level for different companion sizes suggest convergence of our results with respect to $R_c$.

\section{Discussion and Conclusion}
\label{sec:conclusion}
Our 3D PIC simulations nearly approach the scale separation of real spider pulsars. The ratio of companion radius to striped wind wavelength for realistic cases is
\begin{equation}
    \frac{R_{c}}{\lambda}= 100 \left(\frac{R_c}{10^{10} \text{ cm}}\right) \left(\frac{P_s}{3\text{ ms}}\right),
\end{equation} 
while the ratio of the striped wind wavelength to the Larmor radius of particles at $\gamma_\sigma$ is 
\begin{equation}
    \frac{\lambda}{r_{L, \sigma}}=30 \left(\frac{\kappa}{10^4}\right)\left(\frac{d_{\rm IBS}}{10^{11} \text{ cm}}\right)\left(\frac{P_s}{3 \text{ ms}}\right),
\end{equation}
where $\kappa$ is the pair multiplicity at the light cylinder and $d_{\rm IBS}$ is the IBS distance from the pulsar \citep{2025MNRAS.tmp..266C, 2025ApJ...984..146S}.  
In our simulations, $R_c/\lambda=1$ and $\lambda/r_{L, \sigma}\approx30$; while the latter is realistic, the former is far from reality. However, our findings for polarized emission are nearly independent of $R_c/\lambda$ (within the range $0.5\leq R_c/\lambda\leq1$ we explore), 
implying that our predictions are rather robust (see  \citet{2024MNRAS.534.2551C} for the dependence on companion radius in 2D). Future work will need to test whether the same applies in 3D up to realistic ratios $R_c/\lambda\gg1$.

Global 3D PIC modeling enables self-consistent predictions of the inclination-dependent emission and polarization signatures. While 2D modeling captures the main light curve features at $i=90^\circ$, we can now assess the dependence on inclination. We reproduce light curve and polarization patterns predicted in semi-analytic models that assume a toroidal wind incident on a thin IBS \citep{2019ApJ...879...73K, 2023IBSPolarization}. Semi-analytic modeling, however, cannot capture the delicate interplay between particle energization and magnetic field structure. In contrast, all the properties of the emitting particles (energy spectrum, anisotropy, local field geometry) are properly resolved in PIC runs. Thus, our PIC simulations provide novel, first-principles predictions for the PD level in spider pulsars. We find that turbulence caused by field dissipation in the shock downstream suppresses the PD at low $\alpha$, as compared to what is predicted by semi-analytic models \citep{2023IBSPolarization}. Our results can then  improve semi-analytic modeling. Eq. \ref{eq:fit} gives a relationship between PD and $\alpha$ for different inclinations. Thus, observational measurements of the PD will help determine $i$ and $\alpha$.

Before concluding, we discuss a few caveats. We have assumed a plane-parallel pulsar wind where $\alpha$ is the same at all latitudes. In typical black widow systems where the shock wraps around the companion, this assumption is reasonably justified. In these cases, the opening angle of the shock as seen from the pulsar is $R_{\mathrm{curv}}/d_{\mathrm{IBS}}\lesssim0.1$, so substantial variation in $\alpha$ over the IBS would only be expected for small pulsar obliquities \citep{2025MNRAS.tmp..266C}.
In redbacks or black widows where the shock is closer to the pulsar, $\alpha$  would likely vary in latitude over the shock surface. The resulting effects on the shock structure and emission could be probed by future kinetic simulations with $\alpha(z)$. Additionally, in cases where the pulsar spin is not aligned with the orbital angular momentum, $\alpha$ will not be constant along the orbit. For these cases, the $\alpha$ quoted in our work should be considered as the instantaneous value around pulsar superior conjunction, where most of the observed flux (and polarized flux) is produced. 

We predict substantial PD levels for all $\alpha$ and cooling cases. The Imaging X-ray Polarimetry Explorer (IXPE) is capable of significantly detecting PD $\gtrsim15\%$ in the brightest redback systems (with X-ray flux $F_{X}\sim10^{-12}$ erg s$^{-1}$ cm$^{-2}$) with a $\sim1$ Ms exposure \citep{2023IBSPolarization}. Future instruments such as the enhanced X-ray Timing and Polarimetry mission (eXTP) will be able to detect comparable PD levels from more typical redbacks and bright black widows ($F_X\sim10^{-14}-10^{-13}$ erg s$^{-1}$ cm$^{-2}$) with $\lesssim 10$ Ms exposures \citep{2025arXiv250608101Z}. These observations can be directly compared with the results of our kinetic simulations, testing the validity of our models and the properties of the pulsar wind.

\section*{Acknowledgments}
The authors thank Roger W. Romani and Roger D. Blandford for useful discussions. This work was supported in part by a grant from the Simons Foundation (MP-SCMPS-00001470, L.S., A.S.) and facilitated by Multimessenger Plasma Physics Center (MPPC) grant NSF PHY-2206609 to L.S.
A.S. acknowledges the support of the Stanford University Physics Department Fellowship, the National Science Foundation Graduate Research Fellowship, and a Giddings Fellowship at the Kavli Institute for Particle Astrophysics and Cosmology at Stanford. J.C. acknowledges support provided by the NSF MPS-Ascend Postdoctoral Research Fellowship under grant no. AST-2402292. L.S. acknowledges support from DoE Early Career Award DE-SC0023015, NASA ATP 80NSSC24K1238, NASA ATP 80NSSC24K1826, and NSF AST-2307202.

\bibliography{refs}{}
\bibliographystyle{aasjournal}



\end{document}